\documentstyle[12pt]{article}

\title
{
MATTER FIELDS IN THE LAGRANGIAN LOOP REPRESENTATION: SCALAR QED
}

\author{ J.M.~Aroca$^{1}$, M.~Baig$^{2}$, H.~Fort$^{3}$
and R.~Siri$^{4}$ \\
\\
{\sl $^1$Departament de Matem\`atiques,Universitat
Polit\`ecnica de Catalunya,}\\
{\sl 08034 Barcelona, Spain.}\\
{\sl $^2$Grup de F\'{\i}sica Te\`orica, Universitat Aut\`onoma de Barcelona,}\\
{\sl 08193 Bellaterra (Barcelona) Spain.}\\
{\sl $^3$Instituto de F\'{\i}sica, Facultad de Ciencias, Universidad de la}\\
{\sl Rep\'ublica. Trist\'an Narvaja 1674, 11200 Montevideo, Uruguay.}\\
{\sl $^4$Instituto de F\'{\i}sica, Facultad de Ingenier\'{\i}a,
     Universidad de la}\\
{\sl Rep\'ublica. J. Herrera y Reissig 565, 11300 Montevideo, Uruguay.}
}

\date{}

\begin{document}
\bibliographystyle{bibstand}
\maketitle

\begin{abstract}

We present the extension of the Lagrangian $loop$ gauge invariant
representation in such a way to include matter fields. The partition function
of lattice compact $U(1)$-Higgs model is expressed as a sum over closed as much
as open surfaces.  These surfaces correspond to world sheets of loop-like pure
electric flux excitations and open electric flux tubes carrying matter fields
at their ends.  This representation is connected by a duality transformation
with the $topological$ representation of the partition function (in terms of
world sheets of Nielsen-Olesen strings both closed and open connecting pairs of
magnetic monopoles).  We have simulated numerically the $loop$ action
equivalent to the Villain form of the action and mapped out the
$\beta$-$\gamma$ phase diagram of this model. By virtue of the gauge invariance
of this description the equilibrium configurations seems to be reached faster
than with the ordinary gauge-variant descriptions.

\end{abstract}

\newpage

\section{Introduction}

        The use of loops for a  gauge  invariant   description of  Yang-Mills
theories may be traced back to the  Mandelstam \cite{Ma}  quantization without
potentials.   In  1974 Yang \cite{y}   noticed  the  important role of  the
holonomies for a complete description of gauge theories.

     In  the early eighties a loop  based \cite{gt} Hamiltonian  approach  to
quantum electromagnetism was proposed and generalized \cite{gt1} in 1986  to
include the Yang-Mills theory.  This Hamiltonian formulation was given  in
terms of the traces of the  holonomies (the   Wilson  loops)  and their
temporal loop derivatives as the fundamental objects.  They replace the
information furnished by the  vector potential and the electric  field
operator,  respectively.   These gauge invariant operators  verify  a  closed
algebra (non canonical) and  may  be realized  on  a linear space of loop
dependent functions.

        Afterwards the loop representation was extended in such a way to
include matter fields: the so-called $P$-representation \cite{fg},\cite{ggt}.
This extension includes open paths in addition to the closed ones or loops
\footnote{ Here we will use the term `loop' for the configurations in presence
of matter fields albeit in a relaxed sense which covers both closed as much as
open paths.}.

\vspace{3mm}

The loop approach has many appealing features from the methodological point of
view: First,  it allows  to do away with the first class constraints of the
gauge theories (the Gauss law). Second,  the  formalism  only  involves  gauge
invariant objects. This makes it specially  well suited to study `white'
objects  as  mesons  and  barions in QCD because the wave function will only
depend on the paths associated  with the physical excitations.  Third, all the
gauge  invariant  operators have a simple geometrical meaning when realized in
the loop space.  Additionally, there is a conceptual motivation: The
introduction by Ashtekar \cite{a} of a new set of variables that cast general
relativity in the same language as gauge theories allowed to apply loop
techniques as a natural non-perturbative description of Einstein's theory.
Furthermore, the loop representation appeared as the most appealing application
of the loop techniques to this problem \cite{rs}, \cite{gam}. Thus, it was
realized that this loop formalism goes beyond a simple gauge invariant
description and in fact it provides a natural geometrical framework to treat
gauge theories and quantum gravity.

\vspace{3mm}

The non-canonical nature of the loop algebra have made elusive a Lagrangian
loop formalism counterpart of the original Hamiltonian one
\cite{aggs}, \cite{af}.  In a previous paper \cite{af1} we showed a natural
procedure in order to cast loops into the Lagrangian formalism.  In reference
\cite{af1}, it was considered the lattice 4D pure QED model and it was written
the action for the loops as a sum over their closed world sheets.

\vspace{3mm}

In the present paper we shall continue with the program of setting up the
Lagrangian correlative of the Hamiltonian loop formalism. Concretely, we extend
this Lagrangian loop approach in such a way to include matter fields.

We consider the lattice compact version of the scalar electrodynamics
SQED$_c$
which describes the interaction of a compact gauge field $\theta_{\mu}$ with
the scalar field $\phi = |\phi| e^{i\varphi }$.  The self-interaction of the
scalar field is given by the potential $\lambda (|\phi|^2 - \phi_0^2)^2$. For
simplicity we shall consider the limit $\lambda \rightarrow \infty$ which
freezes the radial degree of freedom of the Higgs field (it is known that the
numerical results obtained already at $\lambda=1$ are indistinguishable from
the frozen case). Thus the dynamical variable is compact, i.e. $\varphi \in
(-\pi ,\pi]$.  This model is known to possess three phases, namely confining,
Higgs and Coulomb \cite{j}. The Higgs phase splits into a region where magnetic
flux can penetrate in form of vortices (Nielsen-Olesen strings) and a region
where the magnetic flux is completely expelled \cite{dh}, the relativistic
version of Meissner effect in superconductivity.  Relying on this, we call this
two subregions: Higgs I and II in analogy with superconducting materials.

\section{The Loop Representation: Hamiltonian and Lagrangian
Formulations}

The Hamiltonian loop representation of the scalar compact QED in a lattice --
with sites denoted by $x$, links denoted by $l$ and plaquettes by $p$ -- is
given in terms of the fundamental gauge operators of the P-representation, the
$\hat{\Phi}(P_x^y)$ defined by \footnote{ $P_x^y$ comprises open as much as
closed paths or loops $C$ at fixed $t$, representing "electro-mesons" and pure
gauge excitations respectively.  In this last case $\hat{\Phi}(C)\equiv
\hat{W}(C)$ i.e. it reduces to the Wilson loop operator.}

\begin{equation}
\hat{\Phi}(P_x^y)=\hat{\phi}^{\dagger}(x) \hat{U}(P_x^y) \hat{\phi}(y)
=
\hat{\phi}^{\dagger}(x)\prod_{l\in P} \hat{U}(l) \hat{\phi}(y)
\end{equation}

($\hat{U}(l)$ are the lattice gauge group operators, $\hat{\phi}(x)$ are the
matter field operators) and their conjugate momenta: the electric field
operator $\hat{E}(l)$ and the $\hat{\Pi}^{\dagger}(x)$ and $\hat{\Pi}(x)$.
They  obey the commutation relations

\begin{eqnarray}
\left[ \hat{E}(l), \hat{\Phi}(P_x^y) \right] =
N_l(P) \hat{\Phi}(P_x^y)  \nonumber \\
\left[ \hat{\Pi}^{\dagger}(x), \hat{\Phi}(P_y^z) \right] =
-i{\delta}_{xy} \hat{\Phi}(P_y^z)\nonumber \\
\left[ \hat{\Pi}(x), \hat{\Phi}(P_y^z) \right] =
-i{\delta}_{xz}\hat{\Phi}(P_y^z)
\end{eqnarray}

where $N_l(P)$ is the number of times the link $l$ appears in the path $P$.

The $\hat{\Phi}(P_x^y)$ operators are the creation operators of the loops i.e.

\begin{equation}
\hat{\Phi}(P_x^y)|0>={\hat{\phi}}^{\dagger}(x) {\hat{U}}(P_x^y)
{\hat{\phi}}(y)|0>=|P_x^y>
\label{eq:creation}
\end{equation}

where $\mid 0>$ is the zero loop state (strong coupling vacuum of the system).

\vspace{3mm}

In order to cast the preceding loop description in the Lagrangian formalism let
us consider the partition function for the Villain form of the lattice
action\footnote{ The choice of the Villain form instead of the ordinary Wilson
form is only done for simplicity, with the Wilson action it is also possible to
repeat everything of what follows.} which is given by
\begin{equation}
Z = \int (d\theta ) \sum_{ n }\int (d\varphi ) \sum_{ k } \exp
(-\frac{\beta}{2}\mid\mid \nabla \theta -2\pi n\mid\mid^2
-\frac{\kappa}{2}\mid\mid \nabla \varphi -2\pi k -\theta \mid\mid^2),
\label{eq:Villain}
\end{equation}
where we use the notations of the calculus of differential forms on the lattice
of \cite {g}. In the above expression: $\beta=\frac{1}{e^2}$, $\theta$ is a
real compact 1-form defined in each
link of the lattice and $\varphi$ is a real compact 0-form defined on the sites
of the lattice, $\nabla$ is the co-boundary operator, $n$ are integer 2-forms
defined at the lattice plaquettes, and $k$ integer 1-forms, and $\mid\mid .
\mid\mid^2 = <.,.>$.

        If we use the Poisson summation formula $\sum_n f(n) = \sum_s
\int_{-\infty}^{\infty} d\phi f(\phi) e^{2\pi i\phi s}$ for each of the integer
variables, the partition function (\ref{eq:Villain}) transforms into
\begin{eqnarray}
Z =  \sum_{ s } \sum_{ t } \int (d\theta ) \int (d\varphi )
\int_{-\infty}^{+\infty} (d\psi )
\int_{-\infty}^{+\infty} (d\chi)
\exp(-\frac{\beta}{2}
\mid\mid  \nabla \theta -2\pi \psi \mid\mid^2   \nonumber \\
-\frac{\kappa}{2}
\mid\mid \nabla \varphi -2\pi \chi -\theta \mid\mid^2 )
e^{i2\pi<s,\psi >} e^{i2\pi<t,\chi >}.
\label{eq:Villain1}
\end{eqnarray}

Integrating in the $\psi$ and $\chi$ variables
\begin{eqnarray}
Z \propto  \sum_{ s } \sum_{ t } \int (d\theta ) \int (d\varphi )
\exp(-\frac{1}{2\beta }\mid\mid  s \mid\mid^2
-\frac{1}{2\kappa}\mid\mid \ t \mid\mid^2)  \nonumber \\
\times  e^{i<s,\nabla \theta>} e^{i<t,\nabla \varphi -\theta>}.
\label{eq:V1}
\end{eqnarray}
Using the partial integration rule $<\psi ,\nabla \phi> =<\partial \psi ,\phi>$
($\partial = *\nabla*$ is the boundary operator which maps k-forms into
(k-1)-forms and where $*$ is the duality operation which maps k-forms into
(4-k)-forms) and integrating over the compact $\varphi$ and $\theta$ we get the
constrains $\delta (\partial t = 0)$ and $\delta (\partial s = t)$ and thus, we
finally arrive to
\begin{equation}
Z \propto \sum_{ s } \exp(-\frac{1}{2\beta } \mid\mid  s \mid\mid^2
-\frac{1}{2\kappa }
\mid\mid \partial s \mid\mid^2 )
\label{eq:LOOP}
\end{equation}
or
\begin{eqnarray}
Z \propto \sum_{ s } \exp (  -\frac{1}{2\beta } < s ,\frac{\nabla \partial +
M^2}{M^2} s> ),
\label{eq:LOOP1}
\end{eqnarray}
where $M^2=\frac{\kappa }{\beta }$ is the mass acquired by the gauge field due
to the Higgs mechanism.  If we consider the intersection of one of the surfaces
defined by the integer 2-forms $s$ (open and closed surfaces) with a
$t=\,${\it constant}
plane we get spatial paths or `loops'.  It is easy to show that the creation
operator of those paths is just the the $\hat{\Phi}(P_x^y)$ operator. Repeating
the steps from Eq.(\ref{eq:Villain}) to Eq.(\ref{eq:LOOP}) we get for
$<\hat{\Phi}(P_x^y)>$

\begin{equation}
<\hat{\Phi}(P_x^y)> \propto \sum_{ s } \exp(-\frac{1}{2\beta } \mid\mid  s
\mid\mid^2 -\frac{1}{2\kappa }
\mid\mid \partial s + P_x^y \mid\mid^2 ).
\end{equation}

Thus, we have arrived to an expression of the partition function in terms of
the world sheets of electric string-like configurations: the $loop$
(Lagrangian) representation. In this representation the matter fields are
naturally introduced by means of open surfaces which are the world sheets of
open paths representing the `meson-like' configurations.  In other words, as it
is shown in Fig.1, cutting the above open surfaces with planes $t=\,${\it
constant} we get the corresponding quantum Hamiltonian description in terms of
open paths $P_x^y$ (the $P$-representation).

\section{Duality and the topological representation}

Another equivalent description of the Villain form is the $topological$
representation in terms of the topological objects. As our model has two
compact variables we have two topological excitations: monopoles and
Nielsen-Olesen strings \cite{bpppw}.  The BKT expression for the partition
function of compact scalar QED is obtained via the Banks-Kogut-Myerson
transformation \cite{bkm} (see Appendix) and is given by
\begin{eqnarray}
Z \propto \sum_{n(m)}
\exp
(-2{\pi }^2 \beta <*n(m) ,\frac{M^2}{\partial \nabla + M^2} *n(m)>)
\label{eq:BKT}
\end{eqnarray}
where  $m=\partial *n$ are closed integer 1-forms attached to links which
represent monopole loops and $*n(m)=*n-\partial *q$ are integer 2-forms
attached to plaquettes corresponding to the world sheets of both Dirac and
Nielsen-Olesen strings (with monopole loops as borders).  Thus, comparing
(\ref{eq:LOOP1}) and (\ref{eq:BKT}) we can observe a complete parallelism: in
both representations we have a sum over surfaces, and intersecting with a plane
$t=\,${\it constant} we get closed as much as open strings with point charges
at their ends. In the first case this string-like excitations are `electric'
whilst in the second they are `magnetic'.  Furthermore, there is a duality
transformation connecting the confining and Higgs II sectors of the phase
diagram.  We want to remark that there is a slight difference between both
equivalent descriptions. In the BKT representation monopoles occur at the ends
of both the Nielsen-Olesen strings (physical excitations) and the Dirac strings
(non physical gauge-variant objects) so we have the corresponding two types of
world sheets mixed in the 2-form $*n(m)$ of equation (\ref{eq:BKT}).  On the
other hand the gauge invariant $loop$ description is simpler and completely
transparent from the geometrical point of view.

\section{Numerical Analysis}

        We have performed a standard Metropolis Monte Carlo simulation with the
loop action of (\ref{eq:LOOP}).  We have worked on a hypercubical lattice.
Basic variables are integers $n_{\mu \nu}(r)$ attached to plaquettes $p=(r;\mu,
\nu)$, where  $r$ denotes the site and $\mu,\nu =1-4$ two directions.  These
variables describe the world-sheets of the loop-excitations.

        The loop action of (\ref{eq:LOOP}) is equivalent to the Villain form.
There were not numerical results of the phase diagram (as far as we know) for
the Villain action of SQED$_c$.  Thus, in order to map out the phase diagram,
we made preliminary Monte Carlo runs along lines of fixed $\beta$ or $\gamma$
on a $6^4$ lattice.  Typically, 5,000 iterations were done at each coupling and
measurements of the specific heat were taken. Then the coupling was changed by
$\pm 0.005$,  and another 5,000 iterations were made, etc. The resulting crude
phase diagram (three lines: confining-Coulomb, confining-Higgs and
Higgs-Coulomb) proved very helpful in guiding our larger scale simulations.

        The next step was to localize accuratelly the transition lines in a
$8^4$ lattice. With this aim, we have performed runs of 40,000 iterations after
15,000 termalization sweeps per-point on the plane $\beta-\gamma$.  The phase
diagram is shown in Fig. 2.

        It is worth while to mention that the $loop$ description avoids the
problem of summing over gauge redundant configurations. This is reflected in
the fact that when simulating the $loop$ action the convergence to the
equilibrium configurations seems to be quite faster than when using the
ordinary action in terms of the fields.  Furthermore, we observed the absence
of the strong metastability previously observed in Monte Carlo analysis of the
SQED$_c$ lattice theory with the Wilson action, which makes very difficult the
numerical analysis of the phase transition critical exponents.  This fact is
one of the main advantages for using the action defined in eq. (\ref{eq:LOOP}).

\section{Conclusions}

        First, we have shown how to introduce matter fields in the Lagrangian
$loop$ representation. It turns out that the partition function of gauge-Higgs
theory can be represented as the sum over world sheets of loops (open and
closed).

        Second, a correspondence between the $loop$ and the BKT descriptions is
patent and suggests a duality transformation connecting the confining and the
Higgs II regions of the phase diagram. Perhaps this connection found in the
lattice also holds in the continuum, supporting the claims on the existence of
a new phase in QED \cite{az}.

         Third, simulating numerically a loop-action equivalent to the Villain
form of the SQED model (with frozen radial degree of freedom) we mapped out
its phase diagram.  It turns out that it is very similar to the corresponding
phase diagram for the Wilson form \cite{jersak},
\cite{adg}.
Additionally, these results confirm those previously obtained for the same
model but using the Hamiltonian $loop$ approach \cite{afSQED}, because the
mapped phase diagram is qualitatively equal in both cases.

        The next step will be the implementation of the Pauli exclusion
principle in the Lagrangian loop approach in order to include fermionic fields.
This task has been accomplished in the context of the Hamiltonian $loop$
formalism in reference \cite{fg} where a transparent geometrical description of
`full' QED was given.

\section*{Appendix}

To obtain the monopole representation (\ref{eq:BKT}) we start with the Villain
form (\ref{eq:Villain}) and fix the gauge $\varphi =0$.  Then, we
parameterize the $n=\nabla q + \bar{n}(v)$, where $q$ run over arbitrary
1-forms and $v$ over all co-closed 3-forms ($\nabla v =0$).  $\bar{n}(v)$ is a
solution of $\nabla n=v$.  If we perform a translation $k \rightarrow k-q$
we get the expression

\begin{eqnarray}
Z \propto \int_{-\infty}^{+\infty} (dA) \sum_{
                \begin{array} {c} v\\
                              \left(  \nabla v = 0 \right)
                \end{array}
}\sum_{ k } \exp
(-\frac{\beta}{2}\mid\mid \nabla A-2\pi \bar{n}\mid\mid^2
-\frac{\kappa}{2}\mid\mid A + 2\pi k \mid\mid^2)
\label{eq:Villain_nc}
\end{eqnarray}

where $A=\theta - 2\pi q$ is a non-compact variable.
By shifting
$A$ by $2\pi k$ we find dependence on the combination
$\bar{n}+\nabla k$, which turns to be a solution of
$\nabla n=v$ so we can eliminate the $k$ variable
\begin{eqnarray}
Z \propto \int_{-\infty}^{+\infty} (dA) \sum_{
                \begin{array} {c} v\\
                              \left(  \nabla v = 0 \right)
                \end{array}
}\exp
(-\frac{\beta}{2}\mid\mid \nabla A-2\pi \bar{n}\mid\mid^2
-\frac{\kappa}{2}\mid\mid A \mid\mid^2)
\label{eq:Villain_nc1}
\end{eqnarray}
and performing the
gaussian integration we obtain
\begin{eqnarray}
Z \propto \sum_{
                \begin{array} {c} v\\
                              \left(  \nabla v = 0 \right)
                \end{array}
        } \exp
(-2\pi ^2 \beta <\bar{n}(v),\frac{M^2}{\nabla \partial + M^2}
\bar{n}(v)>)
\label{eq:Monopoles}
\end{eqnarray}

Performing a duality transformation we get (\ref{eq:BKT}) where $m=*v$ are now
integer closed 1-forms ($\partial m= 0$) and $*n(m)=*\bar{n}(v)+\partial *q$ so
$\partial *n=m$.  It is possible to express (\ref{eq:Monopoles}) in terms of
the $k$ instead of the $n$ variables which reflects the presence of Dirac and
Nielsen-Olesen sheets.

\newpage

\section*{Figure captions}

\begin{itemize}

\item Figure 1: Scalar QED surfaces as string propagation.

\item Figure 2: Phase diagram of scalar QED.

\end{itemize}

\vspace{1cm}

\begin{verbatim}
PS files of the figures will be sent by e-mail if
requested to jomaroca@mat.upc.es
\end{verbatim}


\vspace{2cm}

\begin{thebibliography}{99}

\bibitem{Ma} S. Mandelstam,{Ann. Phys. } {\bf 19}, (1962) 1.

\bibitem{y} C.N. Yang, {  Phys. Rev. Lett.}{\bf 33} (1974) 445.

\bibitem{gt}  R. Gambini and A. Trias, {Phys. Rev.} {\bf D22} (1980)
1380.

\bibitem{gt1}  R. Gambini and A. Trias, {Nucl. Phys.} {\bf B278}
(1986) 436

\bibitem{fg} H. Fort and R. Gambini, Phys.Rev. {\bf D44} (1991) 1257.

\bibitem{ggt} R. Gianvittorio, R. Gambini and A. Trias, J.Phys
              {\bf A24} (1991) 3159.

\bibitem{a}  A. Ashtekar, {Phys. Rev. Lett.} {\bf 57}, (1986) 2244;
 {Phys. Rev.} {\bf D36}, (1987) 1587.

\bibitem{rs}  C. Rovelli and L. Smolin, {Phys. Rev. Lett.} {\bf 61}
(1988) 1155; {Nuc. Phys.} {\bf B331} (1990) 80.

\bibitem{gam} R. Gambini {Phys. Lett.} {\bf B235} (1991) 180;

\bibitem{aggs} D. Armand-Ugon, R. Gambini, J. Griego and L. Setaro,
               Preprint IF, Jul.1993.

\bibitem{af} J.M. Aroca and H. Fort, {Phys. Lett.} {\bf B325} (
1994) 166.

\bibitem{af1} J.M. Aroca, M. Baig and H. Fort,
             {Phys. Lett.} {\bf B336} (1994) 54.

\bibitem{j} J. Jersak, in: Lattice gauge theory-a challenge in large
            scale computing, B. Bunk, K.H. Mutter, K. Schilling (eds)
            New York, Plenum Press 1986.

\bibitem{dh} P.H. Damgaard and U.M. Heller, Phys.Rev.Lett. {\bf 60}
             (1988) 1246.

\bibitem{g} A. H. Guth, Phys.Rev.{\bf D21} (1980) 2291.

\bibitem{bpppw} A.K.Bukenov, A.V.Pochinsky, M.I.Polikarpov, L.Polley
                and U.J.Wiese, Nucl. Phys. {\bf B} (Proc.Suppl.)
                {\bf 30} (1993) 723.

\bibitem{bkm}  T. Banks, R. Myerson and J.B. Kogut, Nucl. Phys.
                      {\bf B129} (1977) 493.

\bibitem{az}  M. Awada and R. Zoller, Nucl. Phys. {\bf B365}
(1993) 699 and references therein.

\bibitem{jersak} K. Jansen, J. Jersak, C.B. Lang, T. Neuhaus
and G. Vones, Nucl. Phys. {\bf B265} (1986) 129.

\bibitem{adg} V. Azcoiti, G. Di Carlo and A. F. Grillo,
Phys. Lett. {\bf B258} (1991) 207.

\bibitem{afSQED} J.M. Aroca and H. Fort, {Phys. Lett.}
{\bf B338} (1994) 60.

\end{thebibliography}
\end{document}